\documentclass[twocolumn,showpacs,preprintnumbers,amsmath,amssymb]{revtex4}
\usepackage{epsfig}

\begin{document}
\draft

\title{Return to return point memory}

\author{J. M. Deutsch, Abhishek Dhar and Onuttom Narayan}
\affiliation{
Department of Physics, University of California, Santa Cruz, CA 95064.}

\date{\today}

\begin{abstract}
We describe a new class of systems exhibiting return point memory
(RPM) that are different from those discussed before in the context
of ferromagnets. We show numerically that one dimensional random Ising
antiferromagnets have exact RPM, when configurations evolve from a large
field. However, RPM is violated when started from some stable configurations
at finite field, unlike in the ferromagnetic case.  This implies that the
standard approach to understanding ferromagnetic RPM systems will fail
for this case. We also demonstrate RPM with a set of variables that
keep track of spin flips at each site. Conventional RPM for the spin
configuration is a projection of this result, suggesting that spin flip
variables might be a more fundamental representation of the dynamics. We
also present a mapping that embeds the antiferromagnetic chain in a
two dimensional ferromagnetic model, and prove RPM for spin-exchange
dynamics in the interior of the chain with this mapping.
\end{abstract}


\maketitle
\begin{figure}
\centerline{\epsfxsize=\columnwidth \epsfbox{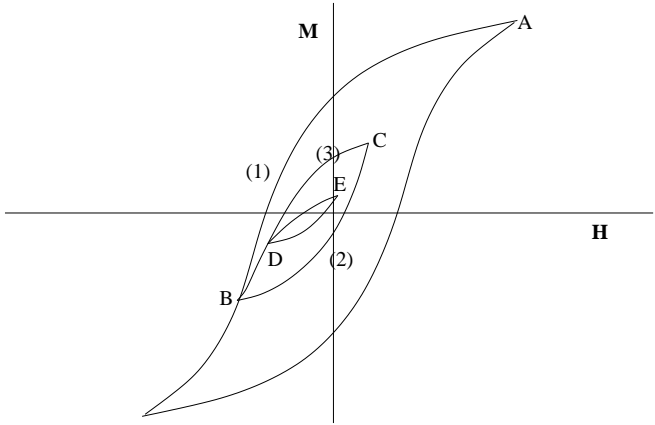}}
\caption{Schematic of hysteresis loop for a ferromagnet, showing return 
point memory (RPM). RPM is seen on the trajectory ABCB, or ABCDED, or 
in general when $H$ backtracks to a value that does not cross the previous
extremum (e.g. E cannot cross C for the path ABCDED to show RPM). For 
zero temperature single spin flip dynamics, it can be proved~\cite{Sethna}
that the full spin configuration on branch (3) is bounded by (1) and (2),
whence RPM follows.}
\label{fig1}
\end{figure}

 
Many systems exhibit the remarkable phenomenon of ``Return Point Memory"
(RPM)~\cite{RPMexpt}. This is most easily demonstrated by looking at
hysteresis loops for ferromagnets, Figure~\ref{fig1}, where an external
field $H$ is lowered from saturation to a minimum value $H_{min}$ and
then raised by some intermediate amount to $H_{max}$ before being lowered
again to $H_{min}$. In systems with RPM, the state of the system the
second time the field reaches $H_{min}$ is identical to the first. The
result generalizes to more complicated variations in $H,$ under the
constraints shown in Figure~\ref{fig1}.

One of the most well known demonstrations of this phenomenon is
Barkhausen noise~\cite{Barkhaus} where the noise observed in changing
$H$ of a ferromagnet is highly reproducible under repeated cycling of
the field~\cite{Barkexpt}.  These experiments show that not only does
the system return to a state having the the same total magnetization
but also gives evidence that the domain wall configurations are identical.

Many systems {\em do not} exhibit RPM. For example spin glass
Hamiltonians with both ferromagnetic and antiferromagnetic bonds,
violate return point memory and instead exhibit subharmonic limit cycles
under the application of a periodic field~\cite{Subharmonic}. Therefore
from a theoretical point of view, it is of interest to try to find
the conditions for a system to exhibit RPM. A major advance in this
direction came from a proof of return point memory for systems with
purely ferromagnetic interactions\cite{Sethna}.  For a broad class of
ferromagnetic models with zero temperature dynamics, RPM was elegantly
proven~\cite{Sethna} by building on the earlier no-passing theorem
for charge density waves~\cite{Middleton}.

The essential ingredient in the proof~\cite{Sethna} of RPM is that
a partial ordering of spin configurations is preserved under the
application of $H$.  By partial ordering, we mean that if one has two spin
configurations $\alpha,\beta$ corresponding to spins $s_1^\alpha,\dots,
s_1^\alpha$ and $s_1^\beta,\dots, s_1^\beta$ then one says that
$\alpha\geq\beta$ if $s_i^\alpha\geq s_i^\beta$ for all $i$. Preservation
of this ordering means that if a time dependent field $H_\alpha(t)$ is
applied to $\alpha$,  and a smaller field $H_\beta(t) \leq H_\alpha(t)$
(for all time $t$) is applied to $\beta$, then $\alpha\geq\beta$ for
all times. That is, $\alpha$ cannot ``pass" $\beta$.  This no-passing
constraint was proved to be valid for ferromagnetic systems with single
spin flip dynamics\cite{Sethna}.

An interesting question to ask is whether or not all systems satisfying
RPM require similar conditions as were needed to prove the ferromagnetic
case\cite{Sethna}. In more detail, is it necessary for {\em every}
pair of two states, that satisfy $\alpha \geq \beta$, to satisfy
the no-passing requirement?  Here we show that, rather surprisingly,
the answer to this question is no.  This is shown by examining
fully antiferromagnetic Ising chains in one dimension, with zero
temperature (deterministic) dynamics identical to those used for
ferromagnetic systems. We find that when started from a large $H$ and
fully saturated magnetization, the system always satisfies RPM. However,
unlike the ferromagnetic case, if one starts in a random state that is
stable at some large field $H_s$, and lowers the field to $H_{min}$,
then raising the field to $H_{max}< H_s$ and returning it to $H_{min}$
changes the state.  From a practical perspective, this is not a severe
restriction, since if $H$ is saturated in the distant past, RPM is valid
for any subsequent evolution of $H(t)$~\cite{foot:temperature}. However,
for the purpose of proving RPM, this implies that there is {\it no\/}
definition of the bounding operator ``$\geq$" (at least none for which
$\alpha\geq\alpha$ is true of all $\alpha$) with which a proof along
the lines of the ferromagnetic case can be constructed.

Even for states descended from the saturated state, for which RPM is
satisfied, we find that no-passing is violated for the spin configuration:
in Figure~\ref{fig1}, the spin configuration on branch (3) is not bounded
above and below by (1) and (2) respectively.  However we have been able
to construct a new ``spin flip" variable that does satisfy no-passing
when starting from a high field. (This is not true
starting from a random configuration; the remarks of the
previous paragraph apply to any variable and bounding operator.) 
No-passing for the spin flip variable implies that it
also satisfies RPM. (However, as with the spin configuration, RPM for a
variable does not imply no-passing.)  Since the spin configuration can
be obtained as a projection of the flip state, this version of RPM is
stronger, suggesting that this new variable may be a more fundamental
way of understanding these systems.

We also investigate whether if one moves beyond single spin flip
dynamics~\cite{Glauber} it is possible for the anti-ferromagnetic chain
to show RPM in the same sense as the ferromagnetic case.  We were able
to show that with spin exchange dynamics\cite{Kawasaki} that, aside from
the ends, conserve magnetization, antiferromagnetic chains do show RPM
in the same configuration-independent way that ferromagnets do. We did
this by embedding this one dimensional antiferromagnetic problem in a
two dimensional {\em ferromagnetic} system which has single spin flip
dynamics and therefore shows return point memory.  However this mapping
must fail for single spin dynamics in a rather interesting way; because
under this mapping, single spin flips become nonlocal and the standard
proof\cite{Sethna} does not apply\cite{foot:itshouldn't} Therefore it
is not just the Hamiltonian, but the dynamics as well, that determine
whether or not a system satisfies RPM.

We consider the random antiferromagnetic Ising model: 
\begin{equation}
{\cal H} = -\sum_i [J_i s_i s_{i + 1} + (h_i + H) s_i]
\label{hamiltonian}
\end{equation}
where the bonds $J_i$ and local fields $h_i$ are independent random
variables.  All the $J_i$'s are negative, and the $h_i$'s are equally
likely to be positive and negative. $H$ is the externally applied
field. Initially, $H$ is large and positive, and all the spins point
up. Thereafter, the field is changed adiabatically. At any field, a spin
is flipped if doing so reduces the energy ${\cal H}$ of the system. This
spin flip can render other spins unstable, in which case the process
is repeated till there are no more spins to flip. If several spins are
unstable, the one whose flipping reduces the energy the most is flipped.
However, because an avalanche propagates outwards from the original
site, and the left and right propagating directions are disjoint,
the same results would be obtained if all unstable spins were flipped
simultaneously. For all the numerical results reported in this paper,
$\sim 10^7$ random choices of $\{J_i, h_i\}$ were tested.

\begin{figure}
\centerline{\epsfxsize=\columnwidth \epsfbox{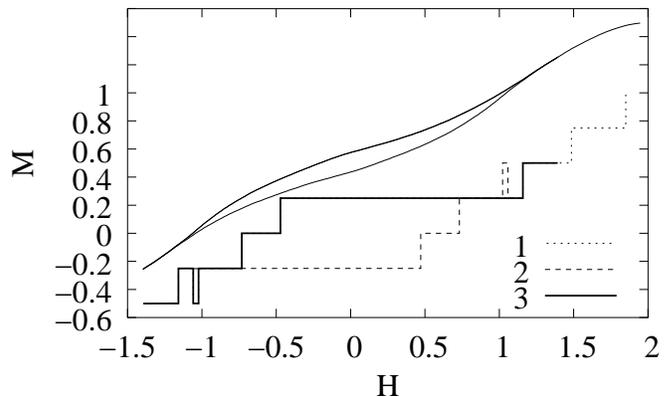}}
\caption{Hysteresis curve for a one-dimensional Ising chain of length
8, where 
the bonds are all antiferromagnetic and of random strength. The 
lower curve shows the hysteresis curve for a single realization of
randomness, with $H$ lowered from $\infty$ to -1.4 (curve 1), 
raised to 1.4 (curve 2) and then lowered to -1.4 again (curve 3). 
The magnetization $M$ changes contrary to $H$ on (2) at 
$H\approx 1.05$ and on (3) at $H\approx -1;$ apart from this 
excursion, (3) coincides with (1) from $-1.4\leq H \leq 1.4.$ The 
upper plot, shifted vertically by $M=0.5$ for clarity, is a similar
graph for a chain of length 64000. Curves (1) and (3)
are different, but so close as to seem indistinguishable in this 
plot. However, return point memory (at $H=-1.4$) is exact.}
\label{fig2}
\end{figure}
Figure~\ref{fig2} shows a typical hysteresis loop, with random bond
disorder but no random fields ($h_i = 0$). The bonds are drawn from
a distribution uniform over $[-1, 0].$ Return point memory is seen at
$H=-1.4$ the hysteresis loop. RPM is also found when the $h_i$'s are drawn
from a distribution uniform over $[-1, 1],$ if $J_i = -1$  for all $i.$
In both cases, although it cannot be shown in the figure, RPM exists for
the full spin configuration rather than just the overall magnetization.
However, if the $J_i$'s are not equal {\it and\/} the random fields are
non-zero, we find that RPM fails if $\delta h \gtrsim 0.01$ and $\delta
J \gtrsim 0.01$ in the random bond and random field cases respectively.
We therefore conclude that either $\delta h$ or $\delta J$ must be
zero for RPM. The results are the same for open and periodic boundary
conditions~\cite{caveat}.  (Even when $\delta h$ and $\delta J$ are both
non-zero, the deviation from RPM is quite small, and hard to detect if
one averages the hysteresis loop over realizations of randomness. A
similar phenomenon was observed earlier for Sherrington-Kirkpatrick
spin glasses~\cite{Pazmandi}.)

An important difference between the ferromagnetic and antiferromagnetic
cases is that a spin at a single site can flip several times while
the magnetic field is varied monotonically. Thus if the field is
lowered, a spin pointing up can be triggered and flip down; if its
neighbors have already flipped down, they can then be pushed back up
by the new spin flip. As a result of this, the magnetization does
not vary monotonically with $H$. This can be seen
in the plot for a single realization of randomness in Figure~\ref{fig2}. 
In more detail,
it is possible observe that i) an avalanche that starts from a
site and destabilizes both its neighbors is only possible for a
configuration $\downarrow\downarrow\uparrow\downarrow\downarrow$ going to
$\downarrow\uparrow\downarrow\uparrow\downarrow$ (or its mirror image),
where the initial site is in the middle. The next nearest neighbors
are stabilized, and the avalanche only covers three sites.  ii) an
avalanche that starts from a site and destabilizes only one neighbor
is only possible for a configuration $\uparrow\downarrow\downarrow$
going to $\downarrow\uparrow\downarrow$ (or its mirror image), where the
initial site is at the end. The avalanche only covers two sites. Thus
as $H$ is varied, the chain evolves through single spin flips, two-site
avalanches with $\Delta M = 0$ and three-site avalanches which have
$\Delta M = 1 $ for decreasing $H$ and $\Delta M = -1$ for increasing
$H.$ These results and more have been proved earlier with random field
disorder (without bond disorder), for the major hysteresis
loop~\cite{Shukla}; the full shape of the hysteresis loop is found
analytically~\cite{Shukla}. We have extended the results of~\cite{Shukla}
to prove ii) for the major loop and i) for the entire hysteresis 
curve~\cite{unpub}.

Motivated by the observation of retrograde variation of the
magnetization with $H,$ we construct an alternative representation of
the dynamics in terms of spin flips instead of the spin configuration.
Initially, when all the spins point up, the flip variable is zero
at each site. Thereafter, each time a spin at site $i$ is reversed,
the flip variable $l_i$ is increased by 1 if this happens when the
field $H$ is increasing, and decreased by 1 if this happens when $H$ is
decreasing. Clearly, along any branch of the hysteresis loop, while $H$
varies monotonically, so must each $l_i.$ Also, $s_i = 1 - 2 [l_i\mod 2],$
and if two configurations $\alpha$ and $\beta$ satisfy the condition that
$l_i^\alpha - l_i^\beta$ is even for all $i,$ they correspond to the same
spin state. In our numerical simulations, we find that for the cases
when RPM is valid, it also holds for the flip configuration. Since the
configuration $\{s_i\}$ is a projection of $\{l_i\},$ this is a stronger
result than RPM, and suggests that the underlying dynamics in terms of
$\{l_i\}$ is fundamental to random antiferromagnetic chains.

With $m_i = \sum_{j\neq i} l_j,$ we find that no-passing is satisfied:
if $H$ is decreased from $H_{max}$ to $H_{min},$ increased to $H_{max}$
and then returned to $H_{min},$ for any $H$ and any site $i$ the value
of $m_i$ on the third segment of this path is bounded below and above
by the corresponding $m_i$'s on the first and second segments (see
Figure~\ref{fig1}).  This is not true for the spin variables 
$s_i$~\cite{foot:li}.
However, as emphasized earlier, RPM is not satisfied if one starts from
an arbitrary initial state at some $H$ instead of the saturated state,
so that unlike the ferromagnetic case~\cite{Sethna} the proof of RPM
must take into account the ancestry of a state.

\begin{figure}
\centerline{\epsfxsize=\columnwidth \epsfbox{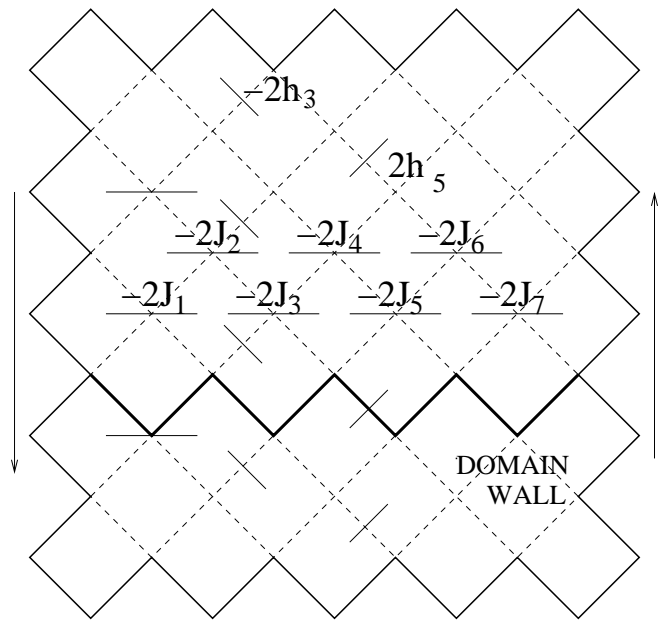}}
\caption{Two dimensional lattice with ferromagnetic bonds. The dashed
lines are to guide the eye; the spins are at the center of each
dashed diamond. The spins are forced to be up and down at the top and
bottom boundaries respectively. The domain wall in between maps to a
one-dimensional spin chain. The case shown corresponds to a chain of
(eight) alternating spins. The horizontal and diagonal bonds in the two
dimensional lattice correspond to the random bonds and fields respectively
of the chain; for the system shown, $h_3<0$ and $h_5> 0$. All bonds
in a vertical column are the same; for clarity, only some are shown.
An external field $H$ on the chain is equivalent to a field at the side
boundaries, increasing as shown with a gradient $H$.}
\label{fig3}
\end{figure}
In order to see to what extent RPM is influenced by the dynamics used 
for the model, we now consider spin-exchange dynamics~\cite{Kawasaki}
instead of single spin flip~\cite{Glauber}. A pair of neighboring 
spins that are oriented opposite to each other are exchanged if it 
is energetically favorable to do so. Since such a move does not change
the overall magnetization, in order for there to be a response to a 
magnetic field, we allow single spin flips at the two ends of the chain
(only open chains are considered). 

This problem can be solved by embedding the antiferromagnetic chain in a
two dimensional ferromagnetic model. We first consider the case when there
is only random bond disorder. Figure~\ref{fig3} shows a two dimensional
square lattice of spins, rotated by an angle $\pi/4.$ Ferromagnetic bonds
connect next nearest neighbors, but (without random fields)
not nearest neighbors. As shown in the figure, the vertical bonds are
all zero, and the horizontal bonds are identical within each vertical
strip. At the top and bottom boundaries, the boundary conditions force
all the spins to be up and down respectively. Free boundary conditions
are used on the side walls. Thus in its ground state, there is one
horizontal domain wall across the system. As shown in the figure, we adopt
a convention in which the domain wall consists of line segments oriented
at $\pm \pi/4,$ i.e. along the principal directions of the square lattice.
The mapping from the two dimensional system to the one-dimensional
chain is as follows: if any line segment of the two dimensional domain
wall is oriented at $\pi/4$ or $-\pi/4,$ the corresponding spin in
the antiferromagnetic chain is 1 or -1 respectively. The two
dimensional ground state corresponds to alternating spins in the chain,
as is appropriate when $H$ is zero.

For a general shape of the domain wall, whenever two successive segments
point in the same direction, a (horizontal) bond is broken, whereas this
does not happen when they point in opposite directions. By choosing the
horizontal bond strengths to be $-2 J_1, -2 J_2, -2 J_3\ldots,$ correlated
vertically, the energy of the antiferromagnetic chain is increased by
$-2J_i$ when spins $i$ and $i+1$ point in the same direction compared
to when they are opposite, as desired for an antiferromagnetic chain.
The magnetic field $H$ couples to $\sum_i s_i$ for the chain, which
is equivalent to the difference in height between the ends of the two
dimensional domain wall. This is equivalent to a magnetic field $H$ on
the rightmost column of the two dimensional system, with the left end of
the domain wall tethered.  It is also possible to generalize the model
to include random bond disorder for the chain: nearest neighbor bonds
of strength $2 |h_i|$ are introduced in the $i$'th column, oriented at
$\pi/4$ if $h_i$ is positive and $-\pi/4$ if $h_i$ is negative.

With this construction, all bonds are ferromagnetic for the two
dimensional system. Further, the fields at the side boundaries vary
monotonically with $H.$ Further, spin-exchange for the chain is equivalent
to single spin flips in the two-dimensional lattice.  The results of
Ref.~\cite{Sethna} can therefore be invoked. We conclude that, with
these dynamics, RPM is valid for all configurations, and is valid for
simultaneous random field and random bond disorder.  As we have seen,
neither of these statements is valid for single spin flip dynamics for
the chain; the two dimensional analog of spin flip at a site on the chain
is to move the entire domain wall to the right of the site up or down by 
one unit if the spin flips up or down~\cite{foot:untether}.

In this paper we have shown that the hysteresis loop for a random Ising 
antiferromagnetic chains at zero temperature exhibits return point 
memory (RPM). For spin flip dynamics, the result is history dependent,
being valid only for configurations that start from saturated magnetization
and a large magnetic field. This is unlike the result for ferromagnets,
where the result is valid for all configurations, indicating that 
the mechanism for RPM is different from the ferromagnetic case. 
(Also, RPM is only valid if either random field or random bond disorder
is present, but not both, a restriction that does not apply to 
ferromagnets.) For spin exchange dynamics, we have proved RPM by 
mapping to a two-dimensional ferromagnetic model, and have therefore
shown that it is as general: valid for all configurations, and with simultaneous
random field and bond disorder. This implies that RPM depends 
on the Hamiltonian and the dynamics used.

We thank Peter Young for useful comments. AD acknowledges support from 
NSF grant DMR 0086287.


\begin{references}
\bibitem{foot:temperature} At non-zero temperature, the spin configuration
will slowly randomize, and could reach an RPM-violating state, whereupon
$H$ would have to be saturated again.
\bibitem{caveat} As $\delta h$ and $\delta J$ are lowered, the
realizations of randomness for which RPM is violated become rarer and
rarer. Therefore, we cannot numerically rule out the possibility that
with simultaneous bond and field disorder there is a non-zero (but less
than 0.01) threshold that has to be crossed before RPM is violated.
\bibitem{foot:itshouldn't} And it should not apply because RPM fails when
evolving from states at finite field as discussed in the text.
\end{references}
\end{document}